# Electron-hole symmetry in a semiconducting carbon nanotube quantum dot

Pablo Jarillo-Herrero†*, Sami Sapmaz†, Cees Dekker†, Leo P. Kouwenhoven†* and Herre S.J. van der Zant†

†*Kavli Institute of Nanoscience Delft and the* *ERATO project on Mesoscopic Correlations, Delft University of Technology, PO Box 5046, 2600 GA, The Netherlands*

**Optical and electronic phenomena in solids arise from the behaviour of electrons and holes (unoccupied states in a filled electron sea). Electron-hole symmetry can often be invoked as a simplifying description, which states that electrons with energy above the Fermi sea behave the same as holes below the Fermi energy. In semiconductors, however, electron-hole symmetry is generally absent since the energy band structure of the conduction band differs from the valence band[1]. Here we report on measurements of the discrete, quantized-energy spectrum of electrons and holes in a semiconducting carbon nanotube[2]. Through a gate, an individual nanotube is filled controllably with a precise number of either electrons or holes, starting from one. The discrete excitation spectrum for a nanotube with *N* holes is strikingly similar to the corresponding spectrum for *N* electrons. This observation of near perfect electron-hole symmetry[3] demonstrates for the first time that a semiconducting nanotube can be free of charged impurities, even in the limit of few-electrons or holes. We furthermore find an anomalously small Zeeman spin splitting and an excitation spectrum indicating strong electron-electron interactions.**

Carbon nanotubes can be metallic or semiconducting depending on their chirality. Electron transport through individual nanotubes has been studied for both classes[2]. Nanotubes of finite length have a discrete energy spectrum. Analogous to studies on semiconducting quantum dots, these discrete states can be filled with electrons, one by one, by means of a voltage applied to a nearby gate electrode[4]. Whereas metallic nanotubes have shown clean quantum dot (QD) behaviour[5-7], this has not been achieved in semiconducting single wall nanotubes (SWNTs). Theory indicates that semiconducting tubes are more susceptible to disorder than metallic ones[8,9]. Disorder typically divides a semiconducting nanotube into multiple islands preventing the formation of a single, well-defined QD. Consequently, the electronic spectrum of semiconducting SWNTs has not been resolved before.

We report here on clean semiconducting tubes and focus on the regime of a few charge carriers (electrons or holes). We use high-purity carbon nanotubes (HiPco[10]), which are deposited with low density on a doped Si substrate (serving as a backgate) that has an insulating $SiO_2$ top layer[11,12]. Individual nanotubes are electrically contacted with source and drain electrodes (50nm Au on 5nm Cr). We then suspend the nanotubes by etching away part of the $SiO_2$ surface[12]. We generally find that removing the nearby oxide reduces the amount of potential fluctuations (i.e. disorder) in the nanotubes, as deduced from transport characteristics.

In this paper we focus on one particular semiconducting device that shows regular single QD behaviour for both few-hole and few-electron doping. The distance between the electrodes in this device is 270nm (Fig. 1a). The dependence of the linear conductance on gate voltage shown in Fig. 1c is typical for semiconducting p and n-type behaviour[13,14]. A low-temperature measurement around zero gate voltage (Fig. 1d) shows a large zero-current gap of about 300meV in bias voltage, reflecting the semiconducting character of this nanotube. The zigzag pattern outside the semiconducting gap is due to Coulomb blockade[4]. These Coulomb blockade features are more evident in Fig. 1e, where a high-resolution measurement of the differential conductance shows the semiconducting gap with the first two adjacent Coulomb blockade diamonds.

The identification of the Coulomb diamonds for the first electron and first hole allows for an unambiguous determination of the particle number as we continue to fill the QD by further changing the gate voltage. Figure 2a shows the filling of holes, one by one, up to 20 holes. The region for the first 2 holes is enlarged in Fig. 2b. The regularity in the Coulomb diamonds indicates a nanotube that is free of disorder. A closer inspection shows that the size of the Coulomb diamonds varies periodically on a smooth background as the hole number increases (Fig. 2c). The alternating, even-odd pattern in this addition energy, $E_{add}$, reflects the subsequent filling of discrete orbital states with two holes of opposite spin[4].

We first focus on the additional discrete lines outside the Coulomb diamonds running parallel to its edges, as for instance indicated by arrows in Fig. 2b. Whereas the upper-left edge of the *N*-hole diamond reflects the ground state energy of the (*N*+1)-hole, the extra lines located at higher voltages, *V*, represent the discrete excitation spectrum for (*N*+1)-holes[4]. The spacing in *V* directly measures the energy separation between the excitations. Such discrete spectra were not obtained before for semiconducting nanotubes.

We now compare the excitation spectra for a particular hole (h) number with the same electron (e) number. The left and right columns in Fig. 3 show the spectra for, respectively, holes and electrons. The upper row compares the spectra for 1h and 1e. The yellow arrows in Fig. 3a point at the first 3 excited states for a single hole. (Note that only lines with positive slopes are observed because of asymmetric tunnel

barriers[4].) Yellow arrows in Fig. 3b indicate the corresponding first 3 excitations for a single electron. (Figure 4 explains this correspondence.) Remarkably, we have simply mirror-imaged the arrows from the hole to the electron side without any adjustment of their spacing. We thus find that the 1h and 1e excitations occur at the same energy. Since one-particle systems are free from particle-particle interactions, this symmetry implies that the confinement potential for electrons is the same as for holes.

Electron-hole symmetry also survives interactions as demonstrated in the lower rows in Fig. 3. Again the arrows pointing at the hole excitations have simply been mirror-imaged to the electron side. Thus, we indeed find that the spectra for 2h and 2e and for 3h and 3e show virtually perfect electron-hole symmetry in the excitation spectra. From a closer look one can see that also the relative intensities of the excitation lines display electron-hole symmetry.

The quality of our data allows for a quantitative analysis. The addition energy is defined as the change in electrochemical potential when adding the ($N$+1) charge to a QD containing already $N$ charges. The constant-interaction (CI) model[4] gives $E_{add} = U + \Delta E$, where $U = e^2/C$ is the charging energy ($C = C_S + C_D + C_G$) and $\Delta E$ is the orbital energy difference between $N$+1 and $N$ particles on the QD. In the case of a semiconductor QD the addition energy for adding the first electron to the conduction band equals $U + E_{gap}$. From the observed gap size of 300meV and $U \approx 50$meV, we determine the semiconducting gap $E_{gap} \approx 250$meV, which corresponds to a nanotube diameter of 2.7nm[3]. AFM measurements, that usually underestimate the real height[15], indicate an apparent tube height of 1.7 ± 0.5 nm.

Since two electrons with opposite spin can occupy a single orbital state, the CI model predicts an alternating value for $E_{add}$, where $E_{add} = U$ for $N$ = odd, and $E_{add} = U + \Delta E$ for $N$ = even. We indeed observe such an even-odd alternation in Fig. 2c with average $\Delta E \approx 4.3$meV throughout the entire range of $N=1$ to $N=30$. Measurements of the Zeeman spin-splitting in a magnetic field (see supplementary information) confirm our assignment of even-odd particle number: Lines corresponding to ground states for odd $N$ split (i.e. total spin = ½), whereas even-$N$ lines do not split (i.e. total spin = 0). Fig. 2e shows the value of the Zeeman energy for the one hole orbital states as a function of magnetic field. The data yield a reduced $g$-factor, $g \sim 1.1$, which is significantly lower than the value $g = 2$ reported on metallic nanotubes[5,7]. (Some experiments on metallic nanotubes report deviations[16].) The reduction in $g$-factor disappears when adding holes. The inset shows that already for 9 holes the normal value is almost recovered. Lower $g$-factors are generally due to spin-orbit coupling, but this effect is small for carbon. It may hint at strong electron-electron interactions in the 1D-QD (see discussion below).

The addition energy spectrum indicates $\Delta E \approx 4.3$meV for consecutive states as we fill the QD with holes. Previous spectra from metallic nanotubes have been analysed by considering a hard-wall potential in the nanotube, with an effective mass determined by the band structure. Our data show that this approach is not justified for semiconducting nanotubes. Lack of effective screening in 1D and the low number of mobile charges yield a gradual potential decay from the contacts[17]. We have computed the addition energy spectrum for a semiconducting nanotube whose gap is ~250meV for two situations (Fig. 2d): hard-wall and harmonic potential of height $E_{gap}/2$ at the contacts[17]. For hard walls the level spacing increases slowly up to ~1.9meV for $N=34$. In the case of a harmonic potential, the level spacing is constant, as in the experiment, and equals 2.7meV, in reasonable agreement with the experimental value ~4.3meV (see supplementary information).

On top of the predicted even-odd pattern, there is a monotonic decrease of the average charging energy with $N$, implying that the total capacitance is changing. We have performed a detailed analysis of the QD electrostatics following ref. [18]. The result is given in the inset to Fig. 2c. It shows that the change in $C$ is mainly due to an increase in $C_S$ and $C_D$. This increase can be assigned to a decrease of the tunnel barrier widths as $|V_G|$ increases, consistent with the simultaneous decrease of d$I$/d$V$ in Fig. 2b. Indeed, d$I$/d$V$ varies from 5GΩ in the first Coulomb peak to 400kΩ at large negative $V_G$.

The observation of electron-hole symmetry poses severe restrictions on the QD system: the effective masses for holes and electrons should be equal and the QD should be free of disorder. Scattering by negatively charged impurities, for example, is repulsive for electrons but attractive for holes, so it would break electron-hole symmetry. A symmetric band structure has been theoretically predicted for graphite materials and carbon nanotubes[3]. In contrast, the absence of scattering has come as a positive surprise.

Fig. 4 clarifies the correspondence between the electron and hole excitation spectra. On the right side of Fig. 4b the situation for electrons is drawn (for $V_G > 0$) and on the left side for holes (for $V_G < 0$). The resulting excitations in transport characteristics as a function of $V$ and $V_G$ then lead to spectra as sketched in Fig. 4c and as measured in Fig. 3.

A detailed analysis of the excitation spectrum requires calculations that are beyond the scope of this paper. The constant-interaction model provides the parameter range for more exact models. The change in orbital energy when adding a charge is given by $\Delta E \approx 4.3$meV, independent of $N$. $\Delta E$ is the scale for the energy difference between single-particle states. Excitations of a smaller energy scale have to be related to interactions. The likely interactions in semiconducting nanotubes are (1) Exchange interaction between spins (e.g. spin = 1 triplet states gain energy from the exchange interaction). Note that we observe an even-odd pattern, which seems to exclude ground states with spins > ½. Excited states, however, can have spins > ½. (2) Electron-phonon interactions. The vibrational modes in a suspended nanotube also have a discrete spectrum, which can show up in the excitation spectra[19]. Note that vibrational modes do not affect the addition energy spectrum of the ground states. (3) Electron-electron interactions. The value for the interaction strength parameter $U/\Delta E \approx 10$. Such a large $U/\Delta E$ ratio points to the presence of phenomena that are not captured by the CI model. Luttinger liquid models developed for finite length metallic nanotubes are not applicable to our few particle nanotubes. A more appropriate starting point are the exact calculations for 1D QDs. In the few particle regime the charge carriers tend to localize and maximize their separation, thereby forming a Wigner crystal[20]. In such a Wigner state, the spectrum consists both of high-energy single particle excitations and collective excitations at low energy[21], similar as in our experiment. Detailed calculations beyond the CI model and a comparison with the experimental results are necessary to establish the precise effect on transport from these interactions.

1. Ashcroft, N. W. & Mermin, N. D. *Solid State Physics* (Saunders College, Orlando, 1976).

2. Dekker, C. Carbon nanotubes as molecular quantum wires. *Physics Today* **52**, 22-28 (1999).

3. Dresselhaus, M. S., Dresselhaus, G. & Eklund, P. C. *Science of Fullerenes and Carbon Nanotubes* (Academic Press, San Diego, 1996).

4. Kouwenhoven, L. P., Austing, D. G. & Tarucha, S. Few-electron quantum dots. *Reports on Progress in Physics* **64**, 701-736 (2001).

5. Tans, S. J. *et al.* Individual single-wall nanotubes as quantum wires. *Nature* **386,** 474-477 (1997).

6. Bockrath, M. *et al.* Single-electron transport in ropes of carbon nanotubes. *Science* **275,** 1922-1925 (1997).

7. Cobden, D. H. & Nygård, J. Shell filling in closed single-wall carbon nanotube quantum dots. *Phys. Rev. Lett.* **89,** 046803-1–046803-4 (2002).

8. Ando, T & Nakanishi, T. Impurity scattering in carbon nanotubes - Absence of back scattering -. *Jpn. J. Appl. Phys.* **67**, 1704-1713 (1998).

9. McEuen, P. L., Bockrath, M., Cobden, D. H., Yoon, Y. & Louie, S. G. Disorder, pseudospins, and backscattering in carbon nanotubes. *Phys. Rev. Lett.* **83**, 5098-5101 (1999).

10. Bronikowski, M. J., Willis, P. A., Colbert, D. T., Smith, K. A. & Smalley, R. E. Gas-phase production of carbon single-walled nanotubes from carbon monoxide via the HiPco process: A parametric study. *J. Vacuum Sci. Technol. A* **19,** 1800-1805 (2001).

11. Fuhrer, M. S. *et al.*, Crossed nanotube junctions, Science **288**, 494-497 (2000).

12. Nygård, J. & Cobden, D. H. Quantum dots in suspended single-wall carbon nanotubes. *Appl. Phys. Lett.* **79,** 4216-4218 (2001).

13. Bachtold, A., Hadley, P., Nakanishi, T. & Dekker, C. Logic circuits with carbon nanotube transistors. *Science* **294,** 1317-1320 (2001) Published online October 4, 2001; 10.1126/science.1065824.

14. Park, J. & McEuen, P. L. Formation of a *p*-type quantum dot at the end of an *n*-type carbon nanotube. *Appl. Phys. Lett.* **79,** 1363-1365 (2001).

15. Postma, H. W. C., Sellmeijer, A., Dekker, C. Manipulation and imaging of individual single-wall carbon nanotubes with an atomic force microscope. *Adv. Mater.* **12,** 1299-1302 (2000).

16. Postma, H.W.C., Yao, Z., and Dekker, C. Electron addition and excitation spectra of individual single-wall carbon nanotubes. *J. Low Temp. Phys.* **118**, 495-507 (2000).

17. S. Heinze *et al*., Carbon nanotubes as Schottky barrier transistors. *Phys. Rev. Lett.* **89**, 106801 (2002).

18. Grabert, H. & Devoret, M. H. *Single Charge Tunneling* (Plenum, New York, 1992).

19. Park, H. *et al.* Nano-mechanical oscillations in a single-$C_{60}$ transistor. *Nature* **407**, 57-60 (2000).

20. Wigner, E. On the interaction of electrons in metals. *Phys. Rev.* **46,** 1002-1011 (1934).

21. Häusler, W. & Kramer, B. Interacting electrons in a one-dimensional quantum dot. *Phys. Rev. B* **47,** 16353-16357 (1993).

Supplementary Information accompanies the paper on www.nature.com/nature

We thank R. E. Smalley and coworkers for providing the high-quality HiPco nanotubes, and S. De Franchesci, J. Kong, K. Williams, Y. Nazarov, H. Postma, S. Lemay and J. Fernández-Rossier for discussions. We acknowledge the technical assistance of R. Schouten, B. van der Enden and M. van Oossanen. Financial support is obtained from FOM.

The authors declare that they have no competing financial interests.

Correspondence and requests for materials should be adressed to P.J. (e-mail: Pablo@qt.tn.tudelft.nl)

**Figure 1** Sample and characterization. **a,** Atomic force microscope image of the device before suspension (scale bar, 200nm). **b,** Device scheme: The nanotube QD is connected to source and drain electrodes via tunnel barriers characterized by resistances $R_S$, $R_D$ and capacitances $C_S$, $C_D$. The backgate is represented by a capacitor $C_G$. The dc source-drain current, $I$, is recorded in the measurements as a function of source-drain voltage $V$ and gate voltage $V_g$. Current-voltage (*I-V*) characteristics are numerically differentiated to obtain the differential conductance, $dI/dV$. **c,** Linear conductance, $G$, as a function of gate voltage, $V_G$, at a temperature, $T \sim 150$ K showing the p and n conducting regions separated by the semiconducting gap. **d,** Large-scale plot of the current (blue, negative; red, positive; white, zero) versus both $V$ and $V_G$ at $T = 4$ K. **e,** High-resolution measurement of the differential conductance as a function of $V$ and $V_G$ in the central region of **d** at 0.3 K. Between $V_G \approx$ -250 and 650 mV, the nanotube QD is depleted entirely from mobile charge carriers. As $V_G$ increases (decreases), one electron (hole) enters the dot as indicated in the right (left) Coulomb diamond.

**Figure 2** Few-hole semiconducting nanotube. **a,** Two-dimensional colour plot of the differential conductance, $dI/dV$, versus $V$ and negative $V_G$ at $T = 4$ K (black is zero, white is 3 µS). In the black diamond-shaped regions the number of holes is fixed by Coulomb blockade. **b,** Zoom in taken at 0.3 K of the region with 0, 1, and 2 holes (white represents $dI/dV > 10$ nS). Lines outside the diamonds running parallel to the edges correspond to discrete energy excitations (the black arrow points at the one electron ground state; the red arrows at the one electron excited states). **c,** Addition energy, $E_{add}$, as a function of hole number. $E_{add}$ is deduced from the diamond size for positive and negative $V$ (i.e. half the sum of the yellow arrows in **a**). Inset, the capacitances $C_S$ (green), $C_D$ (blue) and $C_G$ (black) versus hole number. **d,** Calculation of the addition energy spectrum for a semiconducting nanotube (as an example we have taken a zig-zag (35,0), with $E_{gap} \sim 259$ meV, $m_{eff} = 0.037$ $m_e^3$) for a harmonic potential (top) and a hard-wall potential (bottom). The parameters for the harmonic potential are: $V (x = \pm 135$ nm$) = E_{gap}/2$ (see supplementary information). **e,** Zeeman splitting energy, $E_Z$, versus magnetic field, $B$, for the one hole orbital states. The data result from two different types of measurements: i) individual gate voltage traces at fixed bias (circles) and ii) stability diagrams (squares, see also supplementary information). Inset, $g$-factor as a function of hole number. The point for $N = 1$ is the average of the data in Fig. 2e. The points for $N = 5, 7$ and $9$ are obtained from co-tunnelling (see supplementary information).

**Figure 3** Excitation spectra for different electron and hole numbers demonstrating electron-hole symmetry. $dI/dV$ is plotted versus $(V, V_G)$ at $T = 0.3$ K. **a,** The transition from the 0 to 1h Coulomb diamonds. **b,** Corresponding transition from 0 to 1e. The white dotted lines in **b** are guides to the eye to indicate the diamond edge (not visible for this choice of contrast). **c** and **d,** same for 1-2h and 1-2e. **e** and **f,** Low-bias zoom in of the 1-2h and 1-2e crossings. **g** and **h,** Crossings corresponding to the 2-3h and 2-3e regions. (In **h**, the current switched between two stable positions for positive bias, with corresponding noise in $dI/dV$.)

**Figure 4** Electron-hole symmetry in semiconducting SWNTs. **a,** Band structure (energy $E$ versus wave vector $k$) of a semiconducting nanotube illustrating symmetric valence and conduction bands. Due to quantum confinement, the carriers occupy a set of discrete energy states, shown on the left for hole doping and on the right for electron doping. **b,** Schematic energy diagrams showing transport of holes (left) and electrons (right) across a QD. The levels for the ground state and two excited states for $N = 2$ are drawn. The distance between equivalent levels on the right and left is the same due to equal effective electron and hole masses. The top hole level is accessed from the left for $V > 0$,

whereas the top electron level aligns with the left lead Fermi energy for $V < 0$. The dotted line in the potentials shows the effect of a negatively charged scatterer, which breaks electron-hole symmetry. **c,** Excitation spectra resulting from the energy diagrams in **b**. New levels entering the bias window due to excited states lead to lines in the d$I$/d$V$ plots that run parallel to the diamond edges. (Note that due to asymmetric barriers, only excitation lines with positive slope become visible.) These diagrams can be compared to the experiment in Fig. 3.

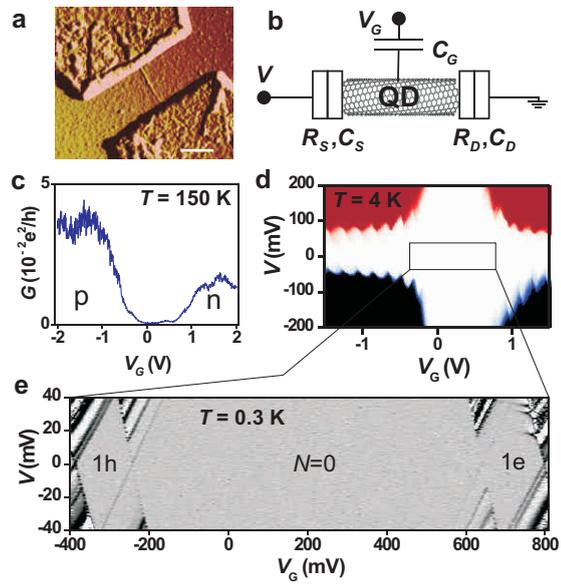

Jarillo-Herrero_fig_1

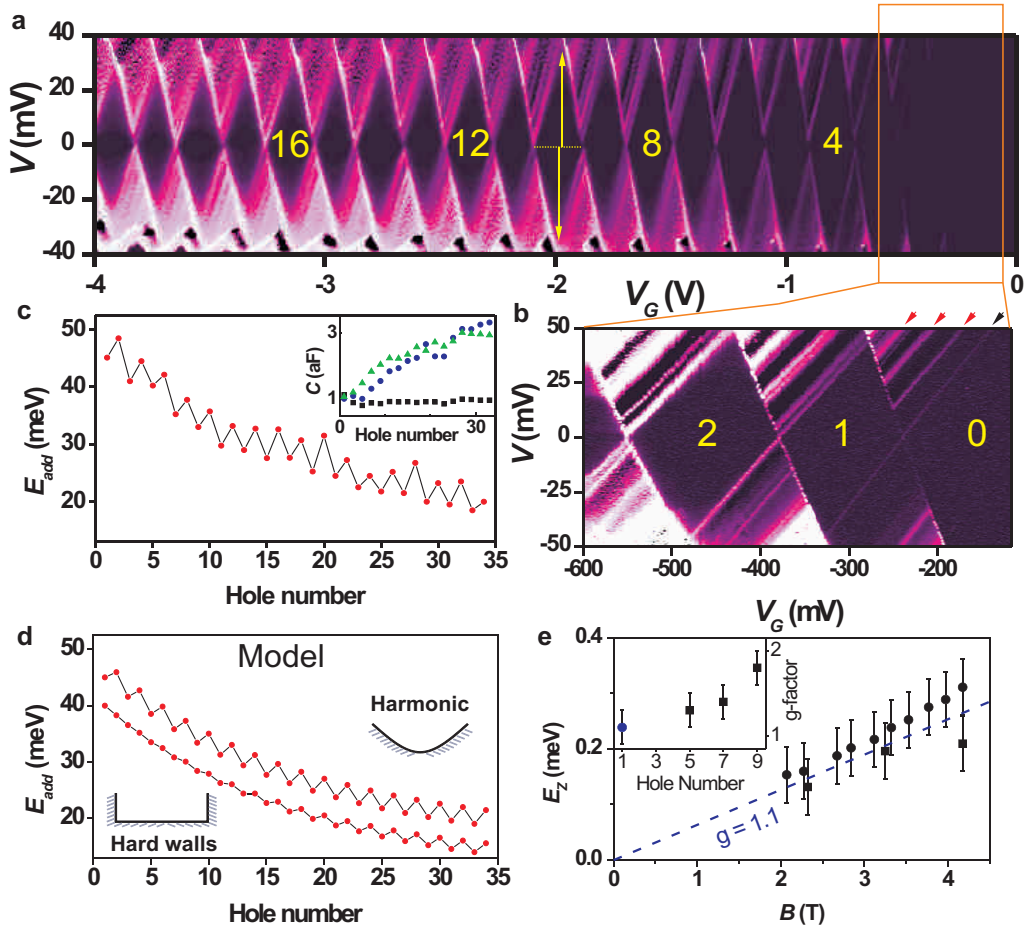

Jarillo-Herrero_fig_2

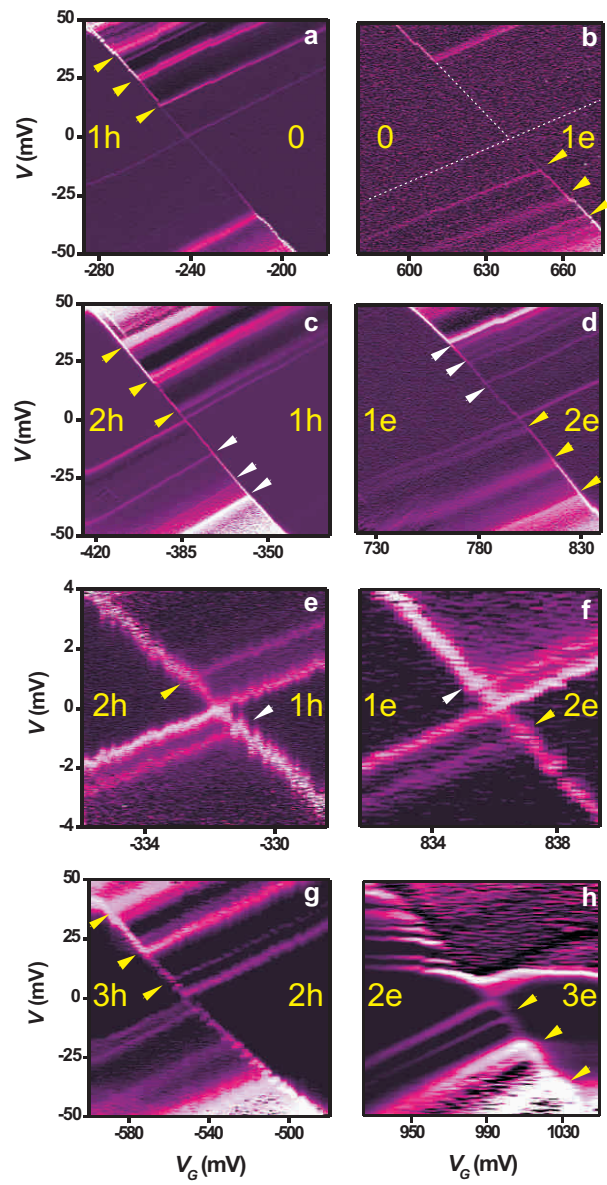

Jarillo-Herrero_fig_3

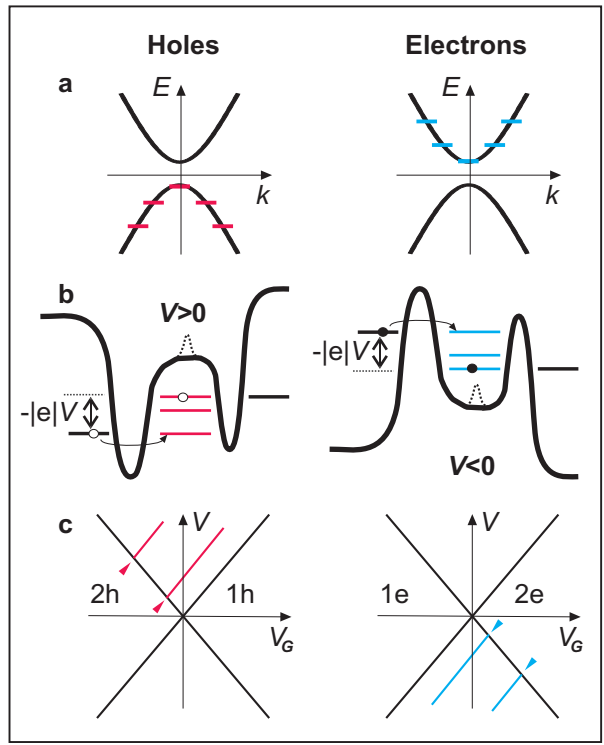

Jarillo-Herrero_fig_4